\documentclass[a4paper,fleqn,usenatbib]{mnras}
\usepackage{multirow}  
\usepackage{psfrag}
\usepackage[T1]{fontenc}
\usepackage[latin9]{inputenc}
\usepackage{float}
\usepackage{amsmath}
\usepackage{amssymb}
\usepackage{graphicx}

\def \x  {x_{l}} 
\def \n  {n_{l}}
\def \taut  {\tilde{\tau}}

\title[Relativistic breakout from a wind]{Relativistic shock breakout from a stellar wind}
\author[Granot, Nakar \& Levinson]{Alon Granot, Ehud Nakar \& Amir Levinson \\
The Raymond and Beverly Sackler School of Physics and Astronomy, Tel Aviv University, Tel Aviv 69978, Israel\\
}

\begin{document}

\maketitle

\begin{abstract}
We construct an analytic model for the breakout of a relativistic radiation mediated shock from 
a stellar wind,  and exploit it to calculate the observational diagnostics of the breakout signal.
The model accounts for photon escape through the finite optical depth wind, and treats the fraction of 
downstream photons escaping to infinity as an adiabatic parameter that evolves in a quasi-steady manner. 
It is shown that the shock is mediated by radiation even when a large fraction of the downstream photons escape,
owing to self-generation and adjustment of opacity through accelerated pair creation.  Relativistic breakout occurs at radii at which the total optical depth of the 
wind ahead of the shock is $\sim (m_e/m_p)\Gamma_{sh}$, provided that the local shock Lorentz factor $\Gamma_{sh}$
exceeds unity at this location.  Otherwise the breakout occurs in the Newtonian regime.  A relativistic breakout is expected in a highly energetic spherical explosion ($10^{52}-10^{53}$ erg) of a Wolf-Rayet star,  or in cases where a smaller  amount of energy ($\sim 10^{51}$ erg)  is deposited by a jet in the outer layers of the star.   
The properties of the emission observed in such explosions during the relativistic breakout are derived. We find that for typical parameters about $10^{48}$ ergs are radiated in the form of MeV gamma-rays over a duration that can range from a fraction of a second to an hour. Such a signal may be detectable 
out to 10-100 Mpc by current gamma-ray satellites.
\end{abstract}

\section{Introduction}
The first light that signals the death of a massive star is emitted upon emergence of the shock 
wave generated by the explosion from the star.
Prior to its breakout the shock propagates in the dense stellar envelope, and is 
mediated by the radiation trapped inside it.   The observational signature of the  breakout event depends on the shock 
velocity and the environmental conditions, thus, detection of the breakout signal and the subsequent emission can provide a wealth 
of information on the progenitor (e.g., mass, radius, mass loss prior to explosion, etc.) and on the explosion mechanism.   

When the shock approaches the edge of the stellar envelope it starts accelerating, owing to the sharp density gradient there. 
Under certain conditions the shock may become mildly and even ultra-relativistic  as it reaches the edge of the star.
For a relatively energetic spherical explosion of $\sim 10^{52}$ erg this happens in compact stars with $R_* \lesssim R_\odot$, while for larger energies, and/or strongly collimated explosions, relativistic shocks are generated also in more extended progenitors \citep{tan2001,nakar2012,nakar2015}.   If the circumstellar 
density is low enough the shock breaks out and undergoes a sudden transition into a collisionless shock at the stellar edge.  However,
if the progenitor is surrounded by a thick enough stellar wind the shock continues to be radiation mediated as it propagates down the wind. The shock physical width then increases during its propagation since the optical depth of the wind decreases, until it reaches the causality scale, $R_{sh}/\Gamma_{sh}^2$, where $R_{sh}$ and $\Gamma_{sh}$ are the shock radius and Lorentz factor, respectively. At this point photons start leaking from the shock to the observer and shock breakout begins. As shown below, in case of relativistic shocks this process may be gradual and can possibly continue over decades in radius. 
This gradual evolution of the shock during  the breakout phase can significantly alter the breakout signal, and 
is of special interest since Wolf-Rayet stars, which are thought to be the progenitors of long GRBs, and are also compact 
enough to have relativistic shock breakout in extremely energetic SNe (such as SN 2002ap),  are known to drive strong stellar winds. 

The breakout of relativistic radiation mediated shocks (RRMS) has been studied recently by \cite{nakar2012}, who employed the infinite planar shock solutions 
obtain by \cite{budnik2010} to show that such episodes give rise to a flash of gamma-rays with very distinctive properties.    However, their analysis is applicable 
to progenitors in which the breakout and the subsequent transition to a collisionless shock  take place at the stellar edge,  
but not to the gradual breakouts anticipated in situations wherein the progenitor is surrounded by a thick stellar wind, as described above.
Shock breakout from a stellar wind has been studied thus far only in the non-relativistic regime, where the shock structure and the physics  underlying 
the breakout process are fundamentally different  \citep[e.g.][]{ofek2010,balberg2011,chevalier2011,svirski2014,svirski2014a} 

The purpose of this paper is to find the observational signature of the breakout of a RRMS from a stellar wind.
The key to that is a proper modeling of the structure of RRMS as photon leakage starts. Here we do that using an analytic approach. 
To be concrete, we suppose that the shock evolves in an adiabatic manner, in the sense that at any given moment its structure can be 
described by a steady solution subject to the local wind conditions.   Upon generalizing the  infinite planar 
shock solutions of \cite{budnik2010} to finite shocks with photon escape, we obtain a family of RRMS solutions which is characterized by one parameter - the fraction of downstream photons escaping the shock to infinity. 
Using these solutions we find the location in the stellar wind where its optical depth cannot support a RRMS anymore and the shock transforms into a collisionless shock. 

Below,  we  first (\S\ref{sec:inf_shock})  describe  and  extend  the analytic  solution  of  the  structure of  a  RRMS
propagating in a medium with and infinite optical depth which  was derived  by \cite{nakar2012}  based on  the numerical
solutions of  \cite{budnik2010}. We note that  this solution is  applicable only to a  RRMS that propagates in  a photon
poor medium, as expected in stellar envelopes and winds.  Under these conditions the photons that mediate the shock must
be  generated within  the shock  transition  layer or  in the  immediate  downstream (see  \citealt{bromberg2011} for  a
discussion on the  relative importance of photons  created in the shock and  photons advected by the  upstream flow). We
then   (\S\ref{sec:escape})   generalize   the   analytic   model   to  incorporate   photon   escape   and   apply   it
(\S\ref{sec:breakout}) to  find the criterion  for shock breakout. In  section \S\ref{sec:observations} we  employ these
results to predict the observational signature of a relativistic  breakout from a wind. We find a closure relation that
the duration,  temperature and energy of  the breakout pulse  must satisfy, and  then proceed to calculate  the observed
emission expected  in a stellar  explosion, as a  function of  the explosion and  progenitor parameters. We  also discuss
briefly the  difference between  a breakout  from a stellar  wind and  a breakout from  a stellar  edge. We  conclude in
\S\ref{sec:summary}.

\section{The structure of an infinite planar RRMS}
\label{sec:inf_shock}
	
\begin{figure}
\includegraphics[width=1\columnwidth]{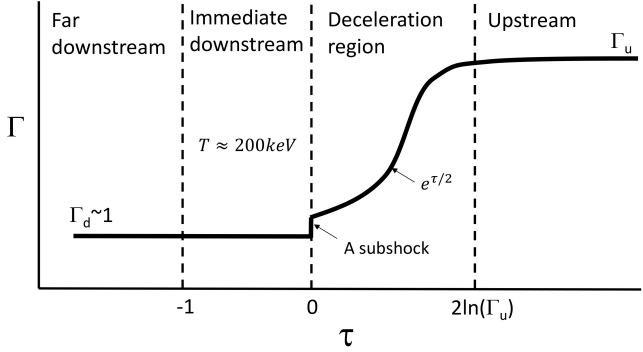}
\caption{Schematic illustration of the Lorentz factor profile, as measured in the shock frame, of an infinite planar RRMS with a photon starved upstream, based on the results of \citealt{budnik2010}. The properties of the different regions indicated in the figure 
are described in the text.}
\label{fig:sketch}
\end{figure}	
	
In this section we outline the analytic solution of a steady, infinite planar RRMS propagating in a photon starved medium. 
This solution was derived by \cite{nakar2012}, based on the numerical solution obtained by \cite{budnik2010}. The main assumption underlying these solutions is that all the plasma components (protons, electrons and pairs) are tightly coupled and behave as a single fluid. This assumption is well justified, since the interaction between the various plasma components occurs over 
scales of the order of the plasma skin depth, which is shorter than the shock width by many orders of magnitude.  

\cite{budnik2010} find that the shock structure can be divided to four regions which are shown 
schematically in figure \ref{fig:sketch}. As seen in the shock frame these regions are:\\
(i) The upstream - unshocked cold plasma moving at Lorentz factor $\Gamma_{u}=\Gamma_{sh}\gg$1.
	The energy density in this region is dominated by the rest mass of the baryons, and is given by $\Gamma_{u}^{2}n_{u}m_{p}c^{2}$,  where $n_u$ is the proton proper density far upstream, $m_p$ is the proton mass and $c$ the speed of light.\\
(ii) The deceleration region - $\Gamma$ is decreasing from $\Gamma_{u}$
	to the downstream value $\Gamma_{d} \approx 1$. The deceleration is driven by the interaction of counter-streaming photons (those moving from the downstream towards the upstream) with the plasma flowing towards the downstream, mostly through  Compton scattering of electron-positron pairs, and pair loading via $\gamma \gamma$ annihilation. \cite{budnik2010} find that at the end of the deceleration zone there is a Newtonian collisionless subshock, which is unimportant for our modeling.\\
(iii) The immediate downstream - the region just downstream of the shock from which the counter-streaming photons that decelerate the relativistic bulk plasma originate. Its optical depth is $\tau \sim 1$. 
The immediate downstream temperature is set by the rate of photon production, mostly through free-free emission, over the available time (roughly the advection time). It equals to $\sim 200$ keV and is largely insensitive to the shock Lorentz factor.  The reason is that in this regime the 
pair density and the resultant photon production rate increase exponentially with increasing temperature, thereby acting as a thermostat that keeps the immediate downstream temperature nearly constant over a large range of shock Lorentz factors.  This regulation mechanism ceases to operate once the temperature exceeds the value above which the dependence of the pair production rate on temperature becomes linear
rather than exponential.  In this regime the analytic model derived below breaks down.  How high the shock Lorentz factor should be for this to occur is unclear at present. The analysis of  \cite{budnik2010} indicates that the model is valid at least up to $\Gamma_{sh} = 30$.   \\
(iv) The far downstream - where the radiation ultimately approches full thermodynamic equilibrium,
and the radiation energy density satisfies $e_r=a_{BB}T_{d}^{4}$, where $T_d$ is the far downstream temperature which for relativistic shocks is
vastly smaller than the immediate downstream temperature. The  photons from this region cannot stream back to the deceleration region and it is, therefore, unimportant for our treatment. 

The analytic model computes the structure of the deceleration region where the flow is still relativistic in the shock frame ($\Gamma \gtrsim 2$). In this region the two-stream approximation can be used.  One stream (the primary beam) consists of the lasma constituents (protons, electrons and pairs) and the back-scattered photons, all of which move towards the downstream.  The counterstream contains photons, each having an energy 
of $\sim m_ec^2$ in the shock frame, that were generated in the immediate downstream and move towards the upstream.  

Since every counterstreaming photon that interacts with the primary beam is converted either to a backscattered photon or
a lepton, it is convenient to treat the pairs created inside the shock and the backscattered 
photons as a single species.  The proper density of this species is the sum $\n = n_{\gamma,\rightarrow d}+n_{\pm}$,
where $n_{\gamma,\rightarrow d}$ and $n_{\pm}$ denote the proper densities of backscattered photons and pairs, respectively.
We use this spices also to define the loading parameter $\x \equiv \frac{\n}{n}$  where $n\equiv n_{p}=n_{e}$ is the proper baryon density
which must be equal to the density of electrons advected from the far upstream (i.e., not created inside the shock) by 
virtue of charge conservation.
The shock equations, derived in the appendix, subject to the boundary condition $x_l=0$ far upstream where counterstreaming photons 
do not reach, then reduce to the following conservation 
laws:  baryon number conservation,
\begin{equation}\label{eq:np}
	n_{u}\Gamma_{u}\beta_{u}=n\Gamma\beta,
\end{equation}
and energy conservation (see equation \ref{app:Gamma-cons}),
\begin{equation}\label{eq:energy}
	\Gamma_{u}=\Gamma(1+4(\x+1)\hat{{T}}\mu),
\end{equation}
where $\mu=\frac{m_e}{m_p}$, $\hat{T}=kT/m_ec^2$ is the dimensionless temperature, $\beta$ is the plasma 3-velocity with respect to the shock 
frame and $\Gamma$ the corresponding Lorentz factor. Henceforth, the subscript $'u'$ refers to quantities at the far upstream. 
The energy equation, Eq.(\ref{eq:energy}), neglects the contribution of the counterstreaming photons to the total energy flux,
which is justified in the region where the primary beam is sufficiently relativistic.

The conservation of photon-and-pair fluxes implies $d(\Gamma \n)=-n'_{\gamma,\rightarrow u} d\tau$, where $n'_{\gamma,\rightarrow u}=\Gamma n_{\gamma,\rightarrow u}$ is the density of 
counterstreaming photons, as measured in the shock frame, and $\tau$  is the net optical depth for conversion of counterstreaming photons
including Compton scattering and pair creation, defined explicitly in equation \ref{app:tpt-opacity}.  Now, in an infinite shock 
counterstreaming photons cannot escape the system. Consequently, at any given location within the shock, for each counterstreaming 
photon there corresponds a quanta of the $\n$ species moving towards the downstream, that is, $n'_{\gamma,\rightarrow u}=\Gamma \n$. 
In terms of $\x = \n/n$ we then obtain: 
\begin{equation}\label{eq:pairs}
	\frac{d\x}{d\tau}=-\x.
\end{equation} 

To find the temperature inside the shock we note that every collision of a counterstreaming photon with the primary beam adds additional
quanta of proper energy $\eta \Gamma m_e c^2$ to the primary beam,  where $\eta$ is a factor of order unity that depends on the
exact energy and angular distributions of pairs and photons inside the shock, as well as other details ignored in the analytic model.
Given that this added energy is shared among the entire plasma constituents, the plasma temperature can be approximated as:
\begin{equation}\label{eq:T}
	\hat{{T}}=\frac{\eta \Gamma\n}{\n + n_e+n_p} = \eta\frac{\Gamma \x}{\x+2}.
\end{equation}

Equations \ref{eq:np}-\ref{eq:T} are applicable in the entire region where $\Gamma \gtrsim 2$, including the far upstream\footnote{Here we account for the electrons that are advected to the shock with the protons from the far upstream, which were neglected by \cite{nakar2012}. We therefore extend their solution, which was applicable only to the deceleration region ,also to the upstream where advected  electrons dominate over created pairs.}.  Thus, their solution provides the structure of the upstream and deceleration zone up to the point that the plasma flow becomes mildly relativistic. This set of equations can be solved analytically. From equation \ref{eq:np} and \ref{eq:energy} we obtain
\begin{equation}\label{eq:G_x}
\begin{split}
	\Gamma(\x)&=&\frac{\sqrt{1+16\mu\Gamma_{u}\eta\frac{\x(\x+1)}{\x+2}}-1}{8\mu\eta\frac{\x(\x+1)}{\x+2}}\\
	&\approx& \left\{ \begin{array}{lr}
						\Gamma_{u} & \x \ll \frac{1}{16\mu\Gamma_{u}\eta}\\
						&\\
						 \sqrt{\frac{\Gamma_u}{4\x\mu\eta}} & \x \gg \frac{1}{16\mu\Gamma_{u}\eta}
					\end{array} \right.
\end{split}
\end{equation}
The value of $\x$ that separates the two regimes marks the transition between the deceleration zone and the upstream. 
Extending the solution to the immediate downstream where $\Gamma\approx 1$ we obtain, using equation \ref{eq:energy}, an approximation for the value of $\x$ in the immediate downstream: $x_0 \approx \frac{\Gamma_u}{4\mu\eta}$. This ignores the energy of counterstreaming photons that may not be 
negligible there. However, it is not expected to alter this result by more than a factor of 2.
Choosing $\tau=0$ at the subshock, which marks the transition between the immediate downstream and the deceleration zone, equation \ref{eq:pairs} 
yields
\begin{equation}\label{eq:x_tau}
	\x=\frac{\Gamma_u}{4\mu\eta} e^{-\tau}.
\end{equation}
Equations \ref{eq:G_x} and \ref{eq:x_tau} implies that in the shock transition layer
\begin{equation}\label{eq:G_tau}
	\begin{array}{rcr}
	\quad\quad\quad\quad\quad\quad &\Gamma(\tau) \approx e^{\tau/2}, &\quad\quad\quad (\Gamma<\Gamma_u).
	\end{array}
\end{equation}
These equations reveal the unique nature of RRMS in a photon poor medium. The shock generates its own opacity via rapid pair production. It is doing so exponentially in the deceleration region within an optical depth of a few from the immediate downstream. The width of the shock, 
measured in terms of the net optical depth of a counterstreaming photon, is $\Delta \tau \approx 2ln(\Gamma_u)$.

To verify the validity and check the accuracy of our model we compare it to the numerical solution derived by \cite{budnik2010}. They present their results using the Thomson optical depth of the shock (including the pairs created within the shock), $\tau_*$. Now, the cross-sections for 
Compton scattering and pair production of a photon heading towards the upstream are both in the KN regime, and are approximately equal (to better
than a factor of 2).
Thus, we can write $d\tau_*=\frac{\sigma_T}{\sigma_{KN}}d\tau$, where  $\sigma_T$ is  the Thomson cross-section and
\begin{equation}
	\sigma_{KN} \approx \frac{3}{8}\cdot\frac{ln(2\Gamma(1+a\hat{{T})})}{\Gamma(1+a\hat{T})}\sigma_{T} .
\end{equation}
Here we use the fact that the energy of counterstreaming photons is $\sim m_e c^2$ with respect to the shock frame, 
and that $\Gamma(1+a\hat{T}) \gg 1$. The factor $`a'$ accounts for the exact angular distributions of the colliding streams, and
is typically of order unity.

\begin{figure}
\includegraphics[width=1\columnwidth]{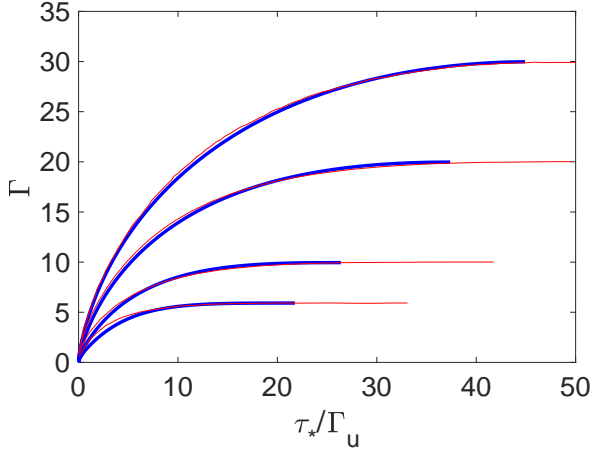}
\caption{A comparison of the Lorentz factor profiles obtained by \citealt{budnik2010} ({\it thin red lines}) and our analytic solution ({\it thick blue lines}), for shocks with $\Gamma_u=$6, 10, 20, 30. Here $\tau_{*}$ is the (pairs loaded) Thomson optical depth. The values of $\eta$ and $a$ are chosen in order to obtain the best fit for each $\Gamma_u$ profile, and are all in the range $\eta=0.45-0.55$ and $a=1.5-2.5$.}
\label{fig:budnikGamma}
\end{figure}

\begin{figure}
\includegraphics[width=1\columnwidth]{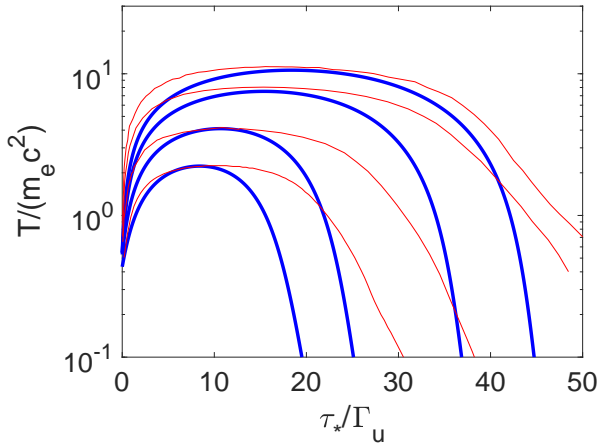}
\caption{Temperature profiles. Notations are the same as in figure \ref{fig:budnikGamma}.}
\label{fig:budnikT}
\end{figure}

\begin{figure}
\includegraphics[width=1\columnwidth]{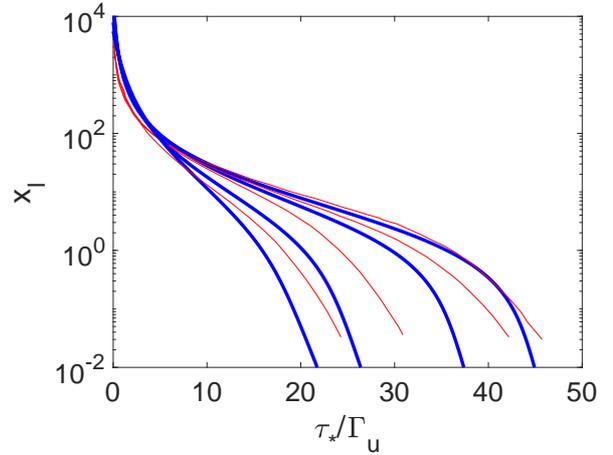}
\caption{Profiles of the loading parameter. The profiles of \citealt{budnik2010} (their figure 10), which are given in terms of positron loading are multiplied by 3, assuming a comparable number of positrons and backscattered photons. Notations are the same as in figure \ref{fig:budnikGamma}.}
\label{fig:budnikX}
\end{figure}

Figure \ref{fig:budnikGamma}-\ref{fig:budnikX} show a comparison of $\Gamma$, $\hat{T}$ and $\x$ profiles as a function $\tau_*/\Gamma_u$ obtained using our analytic model and the numerical model of \cite{budnik2010} (their figures 6, 8 \& 10). The free parameters in this fit are  $\eta$ and $a$ which are expected to be of order unity. We obtain good fits for $\eta=0.45-0.55$ and $a=1.5-2.5$. The agreement is not exact but it is remarkable given the simplicity of the analytic model. In particular, the agreement between the  Lorentz factor profiles and, hence, the shock width is very good. The temperature and pair loading predicted by the analytic model also show general agreement, although there is a significant deviation, especially in the far upstream where our model predicts that pair loading and heating start closer to the shock and grow faster. This is most likely due to our choice of constant values for $\eta$ and $a$ across the entire shock, which corresponds to a constant angular distribution and a constant energy of the counterstreaming photons. However, in reality only the most energetic photons survive in the far upstream due to their smaller cross-section, and this should 
render $\eta$ and $a$ dependent on optical depth.

\begin{figure}
\includegraphics[width=1\columnwidth]{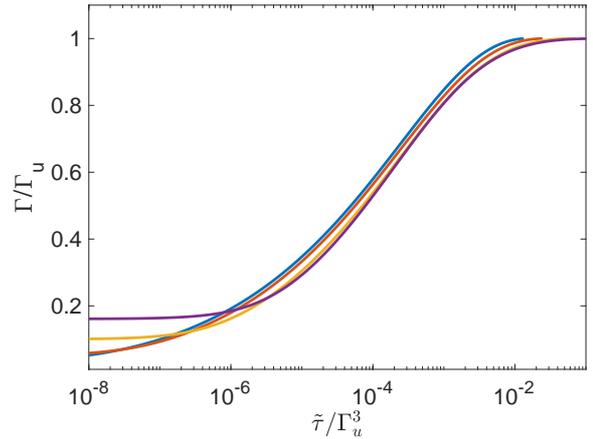}
\caption{Profiles of the  Lorentz factor (normalized to $\Gamma_u$) as a function of $\taut/\Gamma_u^3$, for shocks with $\Gamma_u=$6, 10, 20, 30. Here $\taut$ is the pairs unloaded Thomson optical depth. The values of $\eta$ and $a$ are the same as those used in figure \ref{fig:budnikGamma}. This figure demonstares that all the shocks have similar shapes as a function of $\taut/\Gamma_u^3$ and that their width is consistent with the estimate given in equation \ref{eq:Dtaut}.}
\label{fig:G_taut}
\end{figure}

Finally in order to determine the location at which photons start escaping from a shock propagating in a medium having a finite optical depth, we need to find the width of the shock in terms of the pair unloaded 
Thomson optical depth, $\taut$:
\begin{equation}\label{eq:dtaut}
	d\taut = \frac{d\tau_*}{\x+1}=\frac{\sigma_T}{\sigma_{KN}(\x+1)}d\tau.
\end{equation} 
Note that in a medium with a given density $n$, $d\taut$ is translated directly to a physical distance, $dz=d\taut/(n\sigma_T)$. 
Photon escape will commence once the optical depth to infinity approaches the optical depth of the shock transition layer. In this region $\x \propto \Gamma^{-2}$, $\x \gg 1$ and $\hat{T}\approx\Gamma$, therefore $d\taut \propto \Gamma^4 d\tau$. Since in the deceleration zone $\Gamma \propto e^{\tau/2}$, most of the shock width is contained near the transition from the deceleration region to the upstream, where $\tau$ varies by about one and $\Gamma$ drops from $\Gamma_u$ by about one e-folding. In this region $x \approx \frac{1}{16\mu\Gamma_{u}\eta}$ (see equation \ref{eq:G_x}) and $\Gamma \approx \Gamma_u$. Therefore, from equation \ref{eq:dtaut} we find that the shock width, measured in terms of the pair-unloaded Thomson optical depth, is
\begin{equation}\label{eq:Dtaut}
	\Delta \taut \approx 10\eta\mu\Gamma_{u}^3 \approx \frac{\Gamma_u^3}{400}.
\end{equation}   
Figure \ref{fig:G_taut} exhibits the Lorentz factor profile of the shock as a function of $\taut/\Gamma_u^3$ for several values of $\Gamma_u$. 
It indicates that the profile is universal when plotted as a function of $\taut/\Gamma_u^3$, and that the shock width is consistent with 
equation \ref{eq:Dtaut} in all cases.   We emphasize that this is the main and most important result of our analysis in this section.

It is  interesting to  compare the  optical depth  needed by  the upstream  to sustain  the shock  in the  Newtonian and
relativistic regimes, and its  evolution with the shock 4-velocity, $\Gamma_{sh}\beta_{sh}$. In  the Newtonian regime no
pairs are  created and  there are no  relativistic photons  or electrons  so the relevant  cross-section is  the Thomson
cross-section. Thus, $\taut=\tau$ and $\Delta \taut \approx 1/\beta_{sh}$.  In this regime the shock width is determined
by the requirement that the velocity at which photons  from the downstream diffuse towards the upstream is comparable to
the shock  velocity and thus the  shock becomes narrower  with increasing velocity.  In the relativistic regime,  on the
other hand, there is no diffusion and the width of  the shock is dominated by pair production and Klein-Nishina effects.
Thus, although $\Delta \tau$ depends only logarithmically on  $\Gamma_{sh}$, $\Delta \taut$ (and thus the shock physical
width) depends  sensitively on the shock  Lorentz factor (as $\Gamma_{sh}^3$).  The reason is that  increasing the shock
Lorentz factor reduces both the cross-section and the number of  pairs that are produced at the location where the shock
starts decelerating.  Note, however,  that the  transition in  the physical shock  width between  the Newtonian  and the
relativistic regimes  is not continuous. The  width of a  relativistic shock with  a moderate Lorentz factor  is $\Delta
\taut \ll 1$, much narrower  than the fastest Newtonian shock. This stems from the  vigorous pair production that ensues
once the shock velocity  approches the speed of light. Consequently, the shock  is narrowest when $\Gamma_{sh}\beta_{sh}
\approx 1$, for which $\Delta \taut \sim \mu$, although our solution is not applicable in this regime.

\section{The  structure of  RRMS with  photon escape}\label{sec:escape}  
Next we  explore how  the structure  of a  RRMS
changes as  photons starts  escaping from the  downstream to  infinity, owing  to a finite  upstream optical  depth. The
generalized  model uses  the  same assumptions  invoked in  the  above treatment  of  infinite shocks,  namely a  steady
structure in the shock frame,  and immediate downstream temperature of $\sim 200$ keV. We  discuss the validity of these
two assumptions in the  next section, after presenting the solution.  We also assume that the net  energy carried by the
escaping photons  is subdominant. Under these  assumptions equations \ref{eq:np}, \ref{eq:energy}  and \ref{eq:T} remain
unchanged. The only difference appears when deriving  equation \ref{eq:pairs} since not all the counterstreaming photons
return to  the downstream.  At any location  inside the  shock transition layer  there is a  constant number  density of
photons  (when measured  in  the shock  frame), denoted  by  $n'_{esc}$, that  head  towards the  upstream without  ever
returning (see appendix  for details). Thus, $\Gamma \n=n'_{\gamma,\rightarrow u}-n'_{esc}$  and equation \ref{eq:pairs}
generalizes    to   
\begin{equation}\label{eq:pairs2}    
	\frac{d\x}{d\tau}=-\x-x_{esc}=-\x-f\frac{\Gamma_{u}}{4\eta\mu},
\end{equation}  
where  $x_{esc}=\frac{n'_{esc}}{\Gamma   n}$  and  we  define  $f   \equiv  \frac{x_{esc}}{x_0}  \approx
x_{esc}\frac{4\eta\mu}{\Gamma_{u}}$ to  be the  fraction of downstream  photons that escape  to infinity,  with $0<f<1$,
whereby $f=0$  corresponds to the  limt of no  escape (solved in the  previous section) and  $f\simeq1$ to a  full shock
breakout.

Before proceeding to discuss the  general solution, it is instructive to examine the behaviour  of the solution near the
edge of  a finite  shock of total optical depth to infinity  $\tau_\infty$, and compare  it to  the properties of  a shock  with an
infinite optical  depth at the  location $\tau=\tau_\infty$ (in  both $\tau$ is measured  from the subshock  towards the
upstream). In both shocks the  number of photons that flow towards the upstream  is the same, $n'_{\gamma,\rightarrow u}
(\tau_\infty) = n_0 e^{-\tau_\infty}$. In the finite  shock $n'_{\gamma,\rightarrow u}(\tau_\infty) = n'_{esc}$ while in
the infinite shock  $n'_{\gamma,\rightarrow u}=\n$. Now, in the finite  shock there are no pairs or  photons coming with
the  plasma  at $\tau_\infty$,  namely  $\x(\tau_\infty)=0$,  while  in  the infinite  shock  $\x(\tau_\infty)=x_{esc}$.
Following the  growth of $\x$ in  the finite shock from  the edge towards  the downstream we see  that as long as  $\x <
x_{esc}$ equation  \ref{eq:pairs2} dictates $\x(\tau)\simeq  x_{esc}(\tau_{\infty}-\tau)$. Thus, within an optical
depth of about unity the number  of returning pairs and photons  in the finite shock  becomes comparable to that of  the infinite shock
regardless of the value of  $f$. This is naively expected since, on the average, over  this depth every photon that flow
towards the upstream is  scattered or create a pair. This  implies that since the optical depth of  an infinite shock is
larger than  unity ($\approx 2\ln{\Gamma_{sh}}$),  when the optical  depth is finite and  photons escape, the  shock can
compensate  for that  loss  by producing  enough  returning particles  faster than  in  an infinite  shock,  as long  as
$\tau_\infty \gtrsim 1$. Below we solve the full equations and show that this is indeed the case.

Since equations \ref{eq:np} and \ref{eq:energy} are valid also when photons escape so does equation \ref{eq:G_x} for $\Gamma(\x)$. Combined with equation \ref{eq:pairs2}, in the region where $\x>1$ we obtain
 $d\tau \approx \frac{2\Gamma_{u}-\Gamma}{\Gamma[\Gamma_u(1+f\Gamma^{2})-\Gamma]}d\Gamma$, 
 which can be integrated analytically to yield the shock profile. This expression shows that photon escape becomes important when $f > 1/\Gamma_u^2$. For lower values of $f$ the shock is basically similar to an infinite shock, while for larger values we can approximate
\begin{equation}\label{eq:dtaut_esc}
	d\taut \approx \frac{\mu \Gamma_u}{f} \left[\frac{8a}{3\eta} \frac{G(2-G)}{(1-G)\ln(2a\Gamma_u^2 G)}  \right] dG,
\end{equation} 
where $G \equiv \Gamma/\Gamma_u$. We used the approximated temperature $\hat{T}\approx\eta\Gamma$ which is applicable for $\x>1$ and the relation $x+1=\frac{\Gamma_{u}-\Gamma}{4\eta\mu\Gamma^{2}}$ which stems from equations \ref{eq:energy}. To find the width of the shock, $\Delta \taut $, when photon escape is important we integrate equation \ref{eq:dtaut_esc} from $G=1/\Gamma_u$ to $G=0.9$, which we arbitrarily chose as the boundary between the deceleration zone and the upstream. We find that the integral over the term in the parenthesis depends weakly on $\Gamma_u$ and is of order unity and therefore $\Delta \taut \approx \frac{\mu \Gamma_u}{f}$. Changing the upper limit of the integration to $G=0.99$ changes the result by about a factor of 2. Therefore the shock width in terms of the pair unloaded Thomson optical depth is 
\begin{equation}\label{eq:Dtaut_f}
\Delta\taut=\left\{
\begin{array}{lr}
10\eta\mu\Gamma_{u}^3\hspace{1em}&  f\ll\frac{1}{\Gamma_{u}^{2}} \\
&\\
\frac{\mu \Gamma_{u}}{f}\hspace{1em}& f\gg\frac{1}{\Gamma_{u}^{2}}
\end{array} \right.
\end{equation} 

\begin{figure}
\includegraphics[width=1\columnwidth]{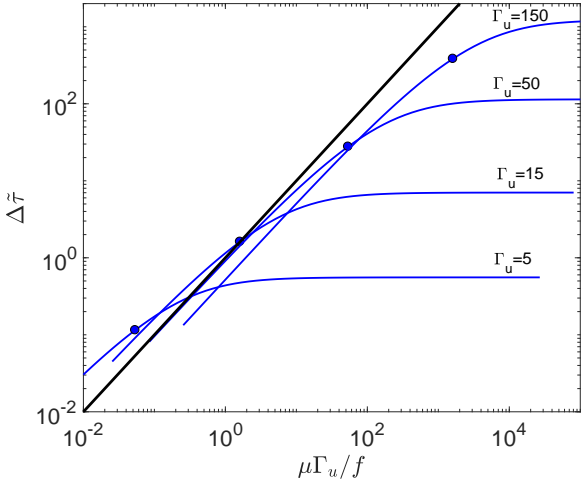}
\caption{The shock width $\Delta \taut$ as a function of $\mu \Gamma_u/f$. On each of the four curves $f$ varies while $\Gamma_u=$5, 15, 50 \& 150 remains constant. $\Delta \taut$ is measured using the full analytic solution with $\eta=0.5$ and $a=2$, and is defined as the pair unloaded Thompson optical depth where $\Gamma=0.9\Gamma_u$. The circles mark the location where $f=1/\Gamma_{u}^{2}$, at which equation \ref{eq:Dtaut_f} predicts a transition in the shock width dependence on $f$. The solid black line is a plot of $\Delta \taut=\mu \Gamma_u/f$, and is given for guidance.  This figure indicates that the width is given by equation \ref{eq:Dtaut_f} to within an order of magnitude.}
\label{fig:Dtaut_f}
\end{figure}
 
\begin{figure}
\includegraphics[width=1\columnwidth]{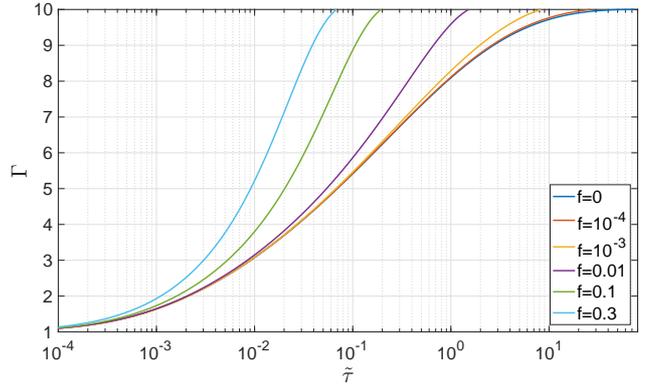}
\caption{Lorentz factor profiles as a function of $\taut$ for shocks with $\Gamma_u=10$ and different $f$ values.}
\label{fig:G_taut_f.eps}
\end{figure}

\begin{figure}
\includegraphics[width=1\columnwidth]{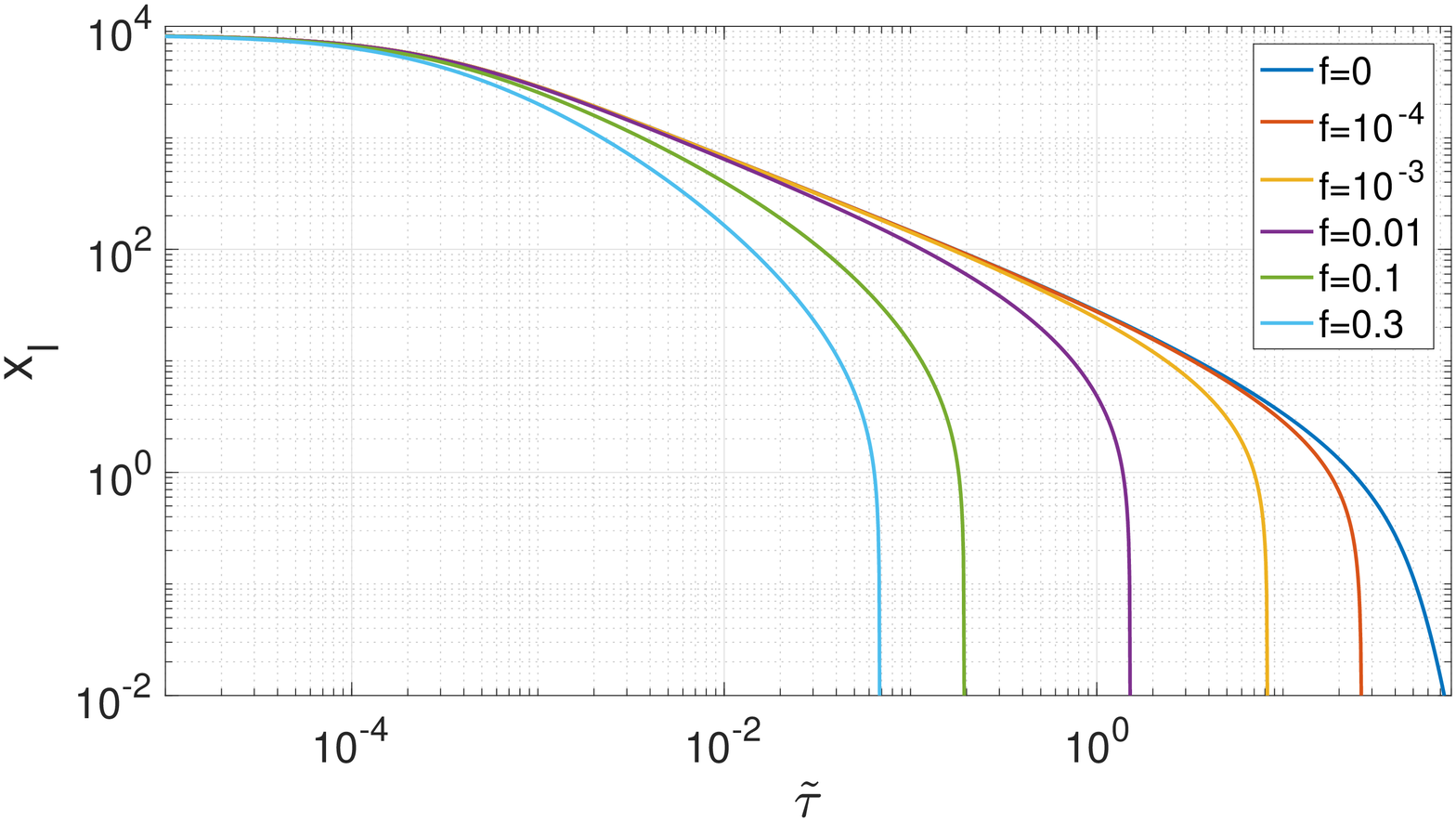}
\caption{Profiles of the  loading parameter as a function of $\taut$ for shocks with $\Gamma_u=10$ and different $f$ values.}
\label{fig:X_taut_f.eps}
\end{figure}

\begin{figure}
\includegraphics[width=1\columnwidth]{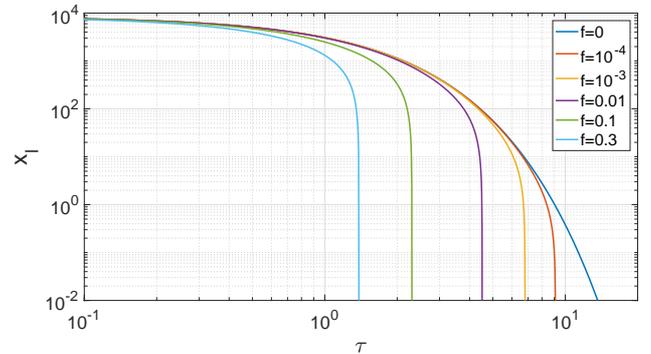}
\caption{Profiles of the  loading parameter  as a function of $\tau$ for shocks with $\Gamma_u=10$ and different $f$ values.}
\label{fig:X_tau_f.eps}
\end{figure}

Figure  \ref{fig:Dtaut_f} shows  $\Delta \taut$,  obtained  from integration  of  the full  equations (without  assuming
$\x>1$), as  a function  $\mu \Gamma_u/f$  for several  values of  $\Gamma_u$ (i.e., on each curve $\Gamma_u$ remains constant while $f$ varies). It  shows that  equation \ref{eq:Dtaut_f}
provides  an order  of magnitude  estimate of  the shock  width  as a  function of  the photon  escape fraction.  Figure
\ref{fig:G_taut_f.eps}  shows the  Lorentz factor  profile as  a function  of $\taut$.  It shows  how the  shock becomes
narrower when  the fraction of  escaping photons increases.  Once the  optical depth to  infinity becomes too  small and
photon escape starts  affecting the shock structure, the shock  adjusts itself and becomes narrower to  match the finite
optical depth. This  is in contrast to a Newtonian  shock which cannot adjust itself, and  therefore full shock breakout
follows    within   a    dynamical    time   after    photon   escape    starts    becoming   non-negligible.    Figures
\ref{fig:X_taut_f.eps}-\ref{fig:X_tau_f.eps} show  the mechanism that  allows the  shock to adjust  its width -  a rapid
buildup of  $\x$ within a unit  interval of the  optical depth $\tau$, which  compensates for the lack  of backscattered
photons  at  $\tau>\tau_{\infty}$.  As  can  be  seen  from  figures  \ref{fig:G_taut_f.eps}-\ref{fig:X_tau_f.eps},  the
structures of all  shocks with different values  of $f$ converge while the  flow is still relativistic,  before it
approaches the  downstream ($\tau \lesssim 1$),  even when photon escape  strongly affects the shock  width. This result
supports our assumption that the conditions in the immediate  downstream are similar in all shocks, even in the presence
of a significant photon escape.


\section{Shock breakout}\label{sec:breakout}  
In the previous section  we derived the  steady state structure of  a RRMS
with  photon escape.  In astrophysical  settings the  shock propagates  in a  medium having  a finite  Thomson depth  to
infinity $\taut_\infty$. Note  that we use $\taut$  to describe the optical  depth of the circumstellar  medium since no
pairs are created before it is encountered by the  shock. For typical settings we expect both $\Gamma_{sh}=\Gamma_u$ and
$\taut_\infty$  to vary  with time.  Then, once  $\taut_\infty \approx  \Delta \taut(f=0)$  photons starts  leaking out.
However, since  the shock width  continuously adjusts, the  breakout is  gradual and the  fraction of shock  energy that
escapes at  every location  can be  estimated roughly  from the  relation $\taut_\infty  \approx \Delta  \taut(f)$, that
yields  $f \approx  \frac{\mu  \Gamma_u}{\taut_\infty}$. A  full  breakout will  take  place once  $f  \sim 1$,  whereby
$\taut_{bo} \sim \mu \Gamma_u$.

Two basic  assumptions underlying the  solution derived in  the previous section;  the first one  is that the  system is
locally in a steady  state, and the second one is  that the immediate downstream temperature is  maintained at $\sim m_e
c^2/3$ throughout the evolution. The former assumption is marginally  justified during a breakout from a wind, since all
the shock properties (e.g.,  $f$ and $\Gamma_{sh}$) vary on the  same timescale as it takes a  photon from the immediate
downstream to cross  the shock and reach the  upstream. Thus, our solution  is a crude approximation, yet  we expect our
estimate of $f$ not  to be strongly affected by the  temporal evolution of the shock structure, as  it merely depends on
the optical depth ahead of the shock; namely, $f$ can  be treated as an adiabatic parameter. The latter assumption, that
the immediate  downstream temperature  is maintained  at $\sim  m_e c^2/3$, holds  provided our  solution for  the shock
structure is stable. We have shown that a self-consistent  solution of the shock structure exists under this assumption,
but whether  it is stable is  yet an open issue.  Thus, although our results  are self-consistent and based  on educated
assumptions they still need to be confirmed by a full time-dependent analysis of the breakout process.

The optical depth to infinity  at the time of full breakout, $\taut_{bo}$, can be  estimated using another method, which
is independent of  the shock structure, except  for the assumption that  the temperature in the  immediate downstream is
$\sim m_e c^2/3$. As  seen in the upstream frame, on  full breakout the shock releases from  the immediate downstream an
energy that is comparable to  the energy that is in the shock transition  layer, namely $E_{bo} \approx\Gamma_u^2 M_{bo}
c^2$  where $M_{bo}$  is the  mass  in the  transition layer,  which is  approximately  the mass  contained between  the
immediate downstream  and infinity. If the  energy is released in  the form of $\sim  \Gamma_u m_e c^2$ photons  (in the
upstream rest frame)  then the total number of  photons upon breakout is $N_{\gamma,bo} \approx  E_{bo}/\Gamma_u m_e c^2
\approx \Gamma_u  M_{bo}/m_e$. Now, for photons  to be able  to escape to  infinity their interaction with  the upstream
electrons should not lead to a runaway of pair production.  Namely, the number of pairs created by this interaction must
be lower than the number  of electrons in the upstream. The breakout will take place  once these numbers are comparable,
namely $N_{\gamma,bo}\tau_{\gamma\gamma} \approx N_e \approx M_{bo}/m_p$. The  energy of photons escaping to infinity is
$\sim \Gamma_u  m_e c^2$ and they  produce pairs on photons  that were scattered once  on electrons in the  upstream and
therefore  their   energy  is  also   $\sim  \Gamma_u  m_e   c^2$,  therefore  $\tau_{\gamma\gamma}   \sim  \taut_\infty
\sigma_{KN}/\sigma_T  \sim  \taut_\infty/\Gamma_u^2  $.  We  therefore   obtain  that  the  breakout  takes  place  once
$\taut_\infty$  satisfies,  \begin{equation}\label{eq:tauBO}  \taut_{bo}   \approx  \mu  \Gamma_{u},  \end{equation}  in
agreement with criterion derived above by setting $f=1$ in equation \ref{eq:Dtaut_f}.

Equation \ref{eq:tauBO} is the  criterion for breakout provided the shock is  relativistic i.e, $\Gamma_{sh}>1$. However
if the  shock decelerates  to mildly relativistic  velocities before  this condition is  satisfied then  the temperature
behind the shock drops to $\sim 50-100$ keV, and the  pair production rate is strongly suppressed. As a result the shock
cannot generate its own opacity any longer and if $\taut_\infty  < 1$ it breaks out. Consequently, in order to determine
the breakout properties in a given astrophysical setting we  need to examine the overall shock evolution and see whether
equation  \ref{eq:tauBO} is  satisfied (i.e.  $f \approx  1$) before  or after  the shock  becomes Newtonian.  If it  is
satisfied before then  the breakout is relativistic, and  takes place once $\taut_{bo} \approx \mu  \Gamma_{u}$. If not,
then the breakout occurs when $\Gamma_{sh} \approx 1$, given that at this time $\taut < 1$.

\section{Observational implications}\label{sec:observations}
In the preceding sections we computed the dependence of the fraction of photons escaping from a RRMS on the optical depth to the observer. Here we consider the observational signature of this process. We first find a closure relation that must be satisfied by the relativistic breakout observables, and then calculate the signature of a RRMS breakout through a stellar wind.

\subsection{Observables as functions of $R_{bo}$ and $\Gamma_{bo}$ and  a closure relation} The three main observables of
a complete  RRMS breakout (i.e.,  $f=1$), the duration,  $t_{bo}$, the observed  temperature $T_{obs,bo}$ and  the total
emitted energy, $E_{bo}$, are determined by only two  phyiscal parameters, the breakout Lorentz factor $\Gamma_{bo}$ and
the  breakout radius  $R_{bo}$: 
\begin{equation}\label{eq:tbo1}  
	t_{bo}\approx \frac{R_{bo}}{2c\Gamma_{bo}^2}  \approx 2 R_{13}  \Gamma_{10}^{-2}  {\rm ~s},  
\end{equation}  
where  $R_{bo}=10^{13}R_{13}$ cm  and  $\Gamma_{bo}=10\Gamma_{10}$,
\begin{equation}\label{eq:Tbo1} 
	T_{obs,bo} \approx 200 \Gamma_{bo} {\rm  ~keV} =2 \Gamma_{10} {\rm ~MeV}, 
\end{equation}
and  
\begin{equation}\label{eq:Ebo1}  
	E_{bo}=3  \times  10^{48} \kappa_{0.2}^{-1}  R_{13}^2  \Gamma_{10}^3  {\rm  ~erg},
\end{equation} 
where $\kappa_{0.2}$  is the Thomson opacity  per unit of mass in  units of $0.2 {~\rm  cm^2/gr}$. In the
last equation  we used the  optical depth of  the shock at  the time of  breakout (equation \ref{eq:tauBO})  and $E_{bo}
\approx \Gamma_{bo}^2  M_{bo} c^2$  where $M_{bo} \approx  4 \pi R_{bo}^2  \taut_{bo}/\kappa$ is  the mass in  the shock
layer upon  breakout. Since  these three observables  depend on two  breakout parameters  they should satisfy  a closure
relation:        
\begin{equation}\label{eq:closure}        
	E_{bo} \approx 10^{48} \kappa_{0.2}^{-1} \left(\frac{t_{bo}}{1{\rm~s}}\right)^2 \left(\frac{T_{obs,bo}}{2{\rm~MeV}}\right)^7 {\rm  ~erg}. 
\end{equation} 
Note the
strong dependence  on the observed  temperature. Since the  expected spectrum  is not a  blackbody this quantity  is not
accurately defined.  This freedom allows  for a wide  range of  breakout observables to  be largely consistent  with the
closure  relation. However,  it does  give some  constraint on  whether  a given  observation may  be a result of a RRMS
breakout. Note also  that the above relations assume  spherical symmetry. However since the breakout  is relativistic it
is enough if  it is quasi-spherically symmetric  over an opening angle that  is larger than $1/\Gamma_{bo}$,  as long as
the observer line-of-sight falls within this opening angle.

\subsection{A  RRMS  breakout  from a  stellar  wind}  
\subsubsection{Spherical  explosion}  
We consider  a  spherically
symmetric explosion of  a star in which  the shock accelerates to relativistic  velocities at the edge  of the envelope,
upon transitioning into  a wind with an  optical depth large enough  to sustain it in a  state of RRMS. As  we will show
this scenario is expected in sufficiently energetic explosions of  Wolf-Rayet (WR) stars. We will also show (in the next
subsection) that the following results may also be applicable, with some adjustments, to aspherical explosions.

The physical  setting that we  consider is as  follows. The explosion  energy (kinetic energy  at infinity) is  $E$, the
progenitor radius  $R_*$ and the  ejecta mass $M_{ej}$.  We approximate  the density at  the stellar edge  as $\rho_*(r)
\propto (R_*-r)^3$, where $r$ is the radius, as expected for a WR star. The wind profile is $\rho_w=A r^{-2}$, where $A$
is a  constant. To find the  observed signature of  a breakout from a  wind in such  settings we find first  the Lorentz
factor of  the shock as  it propagates through  the wind,  as function of  the wind density  and radius, and  then apply
equation \ref{eq:Dtaut_f} to find the emission that escapes to the observer.

Under the  conditions considered  here, the  shock that  was driven by  the explosion  at the  center of  the progenitor
accelerates as it encounters the sharp density gradient near the  stellar edge. For the density profile we use, once the
shock becomes relativistic the  acceleration follows the solution of \cite{johnson1971}  (see also \citealt{tan2001} and
\citealt{pan2006}).  After shock  crossing a  rarefaction wave  crosses the  shocked stellar  material, accelerating  it
farther. The final profile of the expanding  stellar material after its accelleration ends satisfies $E_*(\gamma)\propto
m_*(\gamma) \gamma \propto  \gamma^{-1.1}$, where $E_*(\gamma)$ is  the energy carried by stellar  material with Lorentz
factor $>\gamma$  and $m_*$ is the  corresponding mass \citep{tan2001,barniol-duran2015}.  It is convenient to  find the
normalization  of $E_*(\gamma)$  using the  Lorentz factor  of  $m_*$ that  have an  optical depth  $\tau=1$ before  the
explosion, namely,  $m_{*,1}=4\pi R_*^2/\kappa$.  \cite{nakar2012}  find that\footnote{The  notation here  is different
than in  \cite{nakar2012}. They  use the  symbols $m_0$  for the mass  that have  an optical  depth $\tau=1$  before the
explosion  and $\gamma_{f,0}$  for  its final  Lorentz  factor  after shock  crossing  and expansion.}  \begin{equation}
\gamma_{*,1}  \approx 50  E_{53}^{1.7} M_{ej,5}^{-1.2}  R_{*,11}^{-0.95}, \end{equation}  where $E=10^{53}  E_{53}$ erg,
$M_{ej}=5M_{ej,5} {\rm ~M_\odot}$ and $R_*=10^{11}R_{*,11}$ cm.

After the shock crosses the star it emerges into the wind and starts decelerating. In the process, in addition to the forward shock driven into the wind there is aslo a reverse shock driven into the expanding stellar material. The energy flux through the  reverse shock supports the forward shock an mitigates its deceleration. A rough estimate of the forward shock Lorentz factor as it propagates into the wind can be obtained from equating the energy that crossed the reverse shock with the energy given to the shocked wind\footnote{Note that the shocked wind material has an internal energy of $\sim \Gamma m_p c^2$ per baryon and a bulk Lorentz factor $\sim \Gamma_{sh}$, hence the energy in the shocked wind is $\propto m_w \Gamma_{sh}^2$.}, $m_*(\Gamma_{sh}) \Gamma_{sh} \approx m_w \Gamma_{sh}^2$ where $m_w=4\pi A R_{sh}$ is the wind mass swept by the shock when its radius is $R_{sh}$. Thus, with the equation for $E_*$ we obtain $\Gamma_{sh}=\gamma_{*,1}^{0.68} (m_{*,1}/m_w)^{0.32}$. In order to relate the shock Lorentz factor to the fraction of the shock energy that escapes to the observer, $f$, we use equation \ref{eq:Dtaut_f}, which dictates that once escape starts $\taut_w(R_{sh})=\kappa A/R_{sh}=\mu\Gamma_{sh}/f$. This implies
\begin{equation}\label{eq:Gsh}
	\Gamma_{sh}(f)=\frac{\gamma_{*,1}}{40} \tau_{w,*}^{0.95} f^{-0.48} \approx 1.2 E_{53}^{1.7} M_{ej,5}^{-1.2} R_{*,11}^{-0.95} \tau_{w,*}^{-0.95} f^{-0.48}
\end{equation} 
and
\begin{equation}\label{eq:Rsh}
	R_{sh}(f) \approx 1.7 \times 10^{14} E_{53}^{-1.7} M_{ej,5}^{1.2} R_{*,11}^{1.95} \tau_{w,*}^{1.95} f^{1.48} {\rm~cm}
\end{equation}     
where $\tau_{w,*}=\kappa A/R_* $ is the total optical depth of the wind from the stellar surface to infinity, which is often of the order of unity in WR stars. As the shock propagates $f$ increases while $\Gamma_{sh}$ decreases. The breakout takes place either when $f$ approches unity or when the shock becomes Newtonian and its temperature drops to $\sim 50-100$ keV, so pair production is not efficient enough and the shock cannot generate its own optical depth anymore.

If $\Gamma_{sh}(f=1)>1$ then the breakout is relativistic ($\Gamma_{bo}>1$) and its properties are obtained by setting $f=1$ and plugging equations \ref{eq:Gsh} and \ref{eq:Rsh} into equations \ref{eq:tbo1}-\ref{eq:Ebo1}. The duration of the breakout emission is then
\begin{equation}
	t_{bo}\approx 2000 E_{53}^{-5.1} M_{ej,5}^{3.6} R_{*,11}^{3.85} \tau_{w,*}^{3.86} {\rm ~s}~~~~~(\Gamma_{bo}>1),
\end{equation}
its temperature at $t \sim t_{bo}$ is
\begin{equation}
	T_{obs,bo} \approx 250 E_{53}^{1.7} M_{ej,5}^{-1.2} R_{*,11}^{-0.95} \tau_{w,*}^{-0.95}  {\rm ~keV} ~~~~~(\Gamma_{bo}>1),
\end{equation}
and the total emitted energy is
\begin{equation}
	E_{bo}=10^{48} E_{53}^{1.7} M_{ej,5}^{-1.2} R_{*,11}^{1.05} \tau_{w,*}^{1.05} \kappa_{0.2}^{-1} {\rm ~erg} ~~~~~(\Gamma_{bo}>1).
\end{equation}
The rise time of the breakout emission is much shorter than $t_{bo}$ and from the evolution of emission with $f$ (equations \ref{eq:tbo1}-\ref{eq:Ebo1} and \ref{eq:Gsh}-\ref{eq:Rsh}) we find that during the pulse the luminosity is  almost constant 
\begin{equation}
	L_{bo} \propto t^{0.03}~~~~~(\Gamma_{bo}>1),
\end{equation}
and the temperature drops slowly
\begin{equation}
	T_{obs} \propto t^{-0.2}~~~~~(\Gamma_{bo}>1).
\end{equation} 

If $\Gamma_{sh}(f=1)<1$ then the breakout takes place when the shock becomes mildly relativistic or Newtonian at
\begin{equation}
	R_{bo} \approx R_{sh}(\Gamma_{sh} \approx 1)\approx  3 \times 10^{14} E_{53}^{3.57} M_{ej,5}^{-2.52} R_{*,11}^{-1} \tau_{w,*}^{-1} {\rm~cm} ~~~~~(\Gamma_{bo}\approx 1),
\end{equation}
assuming that $\tau_w<1$ at this location. The duration of the breakout emission is then simply
\begin{equation}
	t_{bo}\approx \frac{R_{bo}}{c} \approx 10^4 E_{53}^{3.57} M_{ej,5}^{-2.52} R_{*,11}^{-1} \tau_{w,*}^{-1}{\rm ~s}~~~~~(\Gamma_{bo}\approx 1),
\end{equation}
and the temperature is
\begin{equation}
	T_{obs,bo} \approx 50-100  {\rm ~keV} ~~~~~(\Gamma_{bo} \approx 1).
\end{equation}
The total emitted energy is roughly $ 4 \pi A R_{bo}c^2$, which depends only on the explosion energy and ejecta mass,
\begin{equation}
	E_{bo} \approx 2 \times 10^{48} E_{53}^{3.57} M_{ej,5}^{-2.52}  \kappa_{0.2}^{-1} {\rm ~erg} ~~~~~(\Gamma_{bo}\approx1).
\end{equation}
while the luminosity depends only on the progenitor radius and wind density
\begin{equation}
   L_{bo} \approx 2 \times 10^{44} R_{*,11} \tau_{w,*}  \kappa_{0.2}^{-1} {\rm ~erg/s} ~~~~~(\Gamma_{bo}\approx1).
\end{equation}

\subsubsection{Aspherical  explosion} 
The  results derived  above are  for spherical  explosions. However,  relativistic
shock breakouts may also occur when  the shock is driven by a relativistic jet, such as  in GRBs and conceivably also in
some SNe  that are not  associated with GRBs  \citep[e.g.,][]{piran2017}. In long  GRBs the jet  successfully penetrates
through the star  and propagates into the  circumstellar medium, driving a  relativistic shock into that  medium. If the
jet's opening angle  lies within the observer's line-of-sight then  the emission from the jet (the  GRB prompt emission)
is expected to outshine  the emission from the breakout of  that shock. However, a relativistic shock  is expected to be
driven into the  cirum-burst medium also away from  the jet opening angle by  the cocoon inflated as the  jet drives its
way through  the star \citep[e.g.,][]{ramirez-ruiz2002,lazzati2010,nakar2017}. In  the case of long  GRBs the properties
of the shock driven by  the emerging cocoon (its opening angle and Lorentz factor) depend  on the mixing between the jet
and the stellar material \citep{nakar2017} and therefore can  be assessed only using numerical simulations \cite[e.g.,][]{gottlieb2017}. We therefore
defer the analysis of the shock breakout through a wind in this case for a future work.

There is strong evidence  \citep[e.g.,][]{bromberg2012} that not in all GRBs the relativistic  jets are powerfull enough
to penetrate  through the  progenitor. Instead,  the jet  dies while  still inside  the star,  leaving a  collimated hot
cocoon. Such  events are called  choked GRBs. There  are also indications  that suggest that  choked jets may  be rather
common in  some types of core-collapse  SNe that are  not associated with GRBs  \citep{piran2017}. The cocoon left  by a
choked jet continues  to propagate through the star and  emerges into the circumstellar medium. Depending  on the energy
deposited by the jet, its opening angle and the depth at which  the jet is choked, the shock driven by the cocoon can be
highly relativistic. in fact  as long as the shock Lorentz  factor is higher than the inverse of  its opening angle, the
spherical theory  derived above  can be applied  for observers that  have their  line-of-sight within the  shock opening
angle. For  these observers the  emission can  be estimated simply  by replacing the  total energy  in the shock  by its
isotropic equivalent energy.  For example, a typical GRB  jet can deposit about $10^{51}-10^{52}$ erg  within an opening
angle of 0.2 rad. If the jet dies after crossing  about half of the progenitor's envelope, then the emerging cocoon will
drive a shock  with an isotropic equivalent energy  of $E_{iso} \sim 5 \times  10^{52}$ - $5 \times 10^{53}$  erg and an
opening angle of about 0.2  rad. The spherical theory is applicable then as a rough  approximation for an observer whose
line-of-sight falls within this opening angle, by replacing $E$ with $E_{iso}$, as long as $\Gamma_{bo} \gtrsim 5$.

Finally, we note that an aspherical breakout from a wind is much simpler to model than an oblique shock breakout from a star. In the latter case the acceleration of the shock leads to a steepening of the shock front angle and the breakout dynamics can become highly nontrivial \citep{matzner2013,salbi2014}. In contrast, a shock that propagates in a wind decelerates and finding its Lorentz factor and shape for a given astrophysical setting is straightforward, although in non-spherical configurations numerical simulations are often needed.

\subsection{The difference between a shock breakout 
from a wind and from a star}\label{sec:windVSstar}
The signature of a relativistic shock breakout from the surface of a star, which was discussed by \cite{nakar2012}, is different than the signature of a breakout from a stellar wind. There are two major differences between the dynamics of shock propagation near the stellar edge and within a stellar wind that reflect on the observed signature. First, near the stellar edge the shock accelerates, while in the wind it decelerates. Second, near the stellar edge the shock velocity varies significantly (from being Newtonian to relativistic) within a very narrow range of radii, while in the wind $\Gamma_{sh}$ varies on the scale of $R_{sh}$. As a result, a shock that propagates near a stellar edge deposits decreasing amounts of energy in an increasingly faster material. After shock breakout all the shocked material expands and the energy deposited by the shock during its acceleration is released. The observer then sees emission that originated from material with a range of Lorentz factors, where  the total emitted energy is not dominated by the radiation released in the shock breakout itself. Instead it is dominated by the energy deposited in material with an optical depth $\taut \approx 1$, namely $m_{*,1}$. The closure relation derived by \cite{nakar2012} for breakout from a star is for the emission released from $m_{*,1}$ after it has reached its final Lorentz factor $\gamma_{*,1}$. In contrast, in case of a wind the shock decelerates and all the internal energy is accumulated between the reverse and forward shocks, where the material has a roughly uniform Lorentz factor, which is comparable to the shock Lorentz factor. The emission is then dominated, at all times, by the leakage of photons from the shock front, and it peaks when $f \approx 1$, at which point the entire energy in the shocked region is released to the observer.

\section{Summary}\label{sec:summary}
In this  paper we  considered a relativistic  shock breakout  from a stellar  wind, assuming that  the evolution  of the
system is sufficiently slow, so  that the structure of the shock can be described, to  a good approximation, by a steady
solution at the local  wind conditions. We further assumed that throughout the  shock evolution the immediate downstream
temperature is  regulated by  vigorous pair production,  and is  maintained at a  level of $\sim  200$ kev.  Under these
assumptions we  constructed an analytic  model for a  finite RRMS  that incorporates photon  losses, and employed  it to
obtain self-consistent  shock solutions  that are characterized  by the  fraction of downstream  photons that  escape to
infinity. Our  model generalizes the analytic  model developed by \cite{nakar2012}  for infinite, planar RRMS,  based on
the numerical solutions of  \cite{budnik2010}. These RRMS are unique in that they  self-generate there own optical depth
via pair production  and, therefore, can propagate also  in a medium at  which the total optical thickness  ahead of the
shock is  much smaller then  unity. The assumption  that the steady  solution provides a  good description of  the shock
structure at  any given time  needs to be  verified with detailed, time  dependent models, as  it is unclear  at present
whether, and under which conditions, the solution we obtained is stable. We leave this problem for a future work.

Our analysis reveals that once photons start leaking through the  upstream of a RRMS, the shock width shrinks by virtue
of accelerated  pair creation that  self-generates the opacity  required to sustain the  shock radiation mediated.  As a
result, the breakout  from a stellar wind is  a gradual process wherein the  fraction of the shock energy  that leaks to
the  observer increases  continuously  over several  decades  in  time and  radius.  This is  in  contrast to  Newtonian
radiation-mediated shocks  in which  the optical depth  is contributed solely  by the  electrons that incident  into the
shock from  far upstream, in which  case the breakout of  the shock and its  transition to a collisionless  shock occurs
within about one dynamical time  after photon leakage commences. A complete breakout of a RRMS  takes place only when it
reaches a radius  beyond which the total optical depth  of the wind becomes $\sim \Gamma_{sh}/1000  \ll 1$. This implies
that in an  explosion of a typical WR  star that is powerful enough to  drive a relativistic shock at  the stellar edge,
shock breakout is  expected to occur from the  wind at a radius much  larger than the stellar radius,  since the optical
depth of a typical WR wind is of order unity.

By combining our analytic shock  solution with existing solutions for the profile of the  ejecta emerging at the stellar
edge, we were  able to predict the observational  signature of a RRMS  breakout through a wind, which  is different than
the signature of a breakout  from a stellar surface (see section \ref{sec:windVSstar}). First  we find the dependence of
the duration,  temperature and  energy of the  breakout emission on  the breakout  radius and Lorentz  factor (equations
\ref{eq:tbo1}-\ref{eq:Ebo1}),  and  show  that  these  three  observables must  satisfy  a  closure  relation  (equation
\ref{eq:closure}). We then find the signal expected from a spherical  explosion of a star. A RRMS in the stellar wind is
generated  if the  explosion is  energetic ($\sim  10^{52}-10^{53}$ erg),  the progenitor  is relatively  compact ($\sim
10^{11}$ cm)  and the  wind optical  depth is  of order  unity. A  similar signal  may be  generated by  a non-spherical
explosion  where  a  choked  jet  deposits  $\sim  10^{51}-10^{52}$  erg within  an  opening  angle  of  $\sim  10$  deg
\citep{piran2017}, if the observer's line-of-sigh falls within the  jet opening angle. For our canonical parameters, the
expected signal is dominated by  emission of gamma-rays ($\sim$MeV) and its energy is  $\sim 10^{48}$ erg. Its duration,
and thus also the luminosity, depends very strongly on the  various parameters and can range from a fraction of a second
to thousands  of seconds. Such  a signal  may be detectable  by gamma-ray observatories  like Swift  and Fermi out  to a
distance of $\sim 10-100$ Mpc.

\section*{Acknowledgements}
We thank Re'em Sari for enlightening discussions. Support by The Israel Science Foundation (grant 1277/13) is acknowledged. AG and EN where partially supported by an ERC starting grant (GRB/SN) and by the I-Core center of excellence of the CHE-ISF.

\appendix
\section{Derivation of shock equations}
We consider a planar shock propagating along the positive $z$ -direction at a velocity $\vec{\beta}=\beta\hat{z}$ (the fluid, in the shock frame, moves in the opposite direction).  
We treat the plasma inside the shock as  a mixed fluid consisting of protons, electrons, positrons and photons in thermal equilibrium.  
For convenience we distinguish between the electrons advected into the shock by the upstream flow and the 
pairs produced inside the shock through $\gamma\gamma$ annihilation.   The proper density of the former is 
denoted by $n_e$ and of the latter by $n_\pm$.    Charge conservation implies $n_e=n$ and $n_-=n_+$, where
$n$ denotes the proper baryon density.    Baryon number conservation,
\begin{equation}
\partial_\mu(nu^\mu)=0,
\end{equation}
here $u^\mu=\Gamma(1,-\beta,0,0)$ denotes the bulk 4-velocity of the mixed fluid with respect to the shock frame, 
implies that $\Gamma n$ is conserved in the relativistic limit $\beta\simeq1$,
and must equal the far upstream value, viz., $\Gamma n =\Gamma_u n_u$.
We further denote the proper density of photons streaming with the flow (i.e., moving from 
the upstream to the downstream) by $n_{\gamma\rightarrow d}$ and the proper density of counterstreaming photons by $n_{\gamma\rightarrow u}$.
The counterstreaming photons  are inverse Compton scattered by the inflowing electrons and positrons, and are annihilated  via interactions
with scattered photons that are moving with the bulk flow.   We suppose that the bulk flow and the counter-streaming photons are both highly beamed 
and adopt the two-beam approximation, that hold in the region where the flow is sufficiently relativistic.  The change in the number density  of counterstreaming
photons is then governed by the equation
\begin{equation}
\frac{dn^\prime_{\gamma\rightarrow u}}{dz}=-[\sigma_{KN} (n^\prime_\pm+n^\prime_e)+\sigma_{\gamma\gamma}n^\prime_{\gamma\rightarrow d}]n^\prime_{\gamma\rightarrow u},
\label{app:rate1}
\end{equation}
where superscript "prime" refers to quantities measured in the shock frame, that is, $n^\prime_e=\Gamma n_e,  n^\prime_\pm=\Gamma n_\pm$, etc., and $\sigma_{KN}$, $\sigma_{\gamma\gamma}$ are the full cross-sections for Compton scattering and pair-production, respectively.
The change in the density of pairs and downstream moving photons are likewise given by
\begin{equation}
\frac{dn^\prime_\pm}{dz}= -2\sigma_{\gamma\gamma}n^\prime_{\gamma\rightarrow d}n^\prime_{\gamma\rightarrow u},
\label{app:rate2}
\end{equation}
and
\begin{equation}
\frac{dn^\prime_{\gamma\rightarrow d}}{dz}=-[\sigma_{KN} (n^\prime_\pm+n^\prime_e)-\sigma_{\gamma\gamma}n^\prime_{\gamma\rightarrow d}]n^\prime_{\gamma\rightarrow u}.
\label{app:rate3}
\end{equation}
The net density of quanta (pairs + photons) produced inside the shock via conversion of counterstreaming photons is given by $\n=n_\pm+n_{\gamma\rightarrow d}$.  Combining Eqs. (\ref{app:rate2})
and (\ref{app:rate3}) yields 
\begin{equation}
\frac{dn^\prime_{l}}{dz}=-[\sigma_{KN} (n^\prime_\pm+n^\prime_e)+\sigma_{\gamma\gamma}n^\prime_{\gamma\rightarrow d}]n^\prime_{\gamma\rightarrow u}.
\label{app:ratel}
\end{equation}
Subtracting equation \ref{app:ratel} from equation  \ref{app:rate1}	  we readily obtain $d(n^\prime_{\gamma\rightarrow u}-n^\prime_l)/dz=0$.   Using the far upstream boundary condition $n^\prime_l(z\rightarrow\infty)=0$
then gives 
\begin{equation}
n^\prime_{\gamma\rightarrow u} = n^\prime_l + n^\prime_{esc},
\end{equation}
where $n^\prime_{esc}=n^\prime_{\gamma\rightarrow u}(z\rightarrow\infty)$ denotes the number density of counterstreaming photons that escape the shock to infinity.
The last equation simply means that every photon that moves from the immediate downstream  towards the upstream either escape the shock or eventually comes back
in the form of a scattered photon or a member of an electron-positron pair.

It is convenient to use the net optical depth for conversion of counterstreaming photons
\begin{equation}
d\tau=[\sigma_{KN}(n^\prime_\pm+n^\prime_e)+\sigma_{\gamma\gamma} n^\prime_{\gamma\rightarrow d} ]dz,
\label{app:tpt-opacity}
\end{equation}
and the fractions $\x=n^\prime_l/n^\prime$, $x_{esc}=n^\prime_{esc}/n^\prime$, as free variables.   The above rate equations
then reduce to the single equation
\begin{eqnarray}
\frac{d \x}{d\tau}= -(\x+x_{esc}).
\end{eqnarray}

The energy momentum tensor of the mixed fluid is the sum of the different components,
$T^{\mu\nu}=T^{\mu\nu}_b+T^{\mu\nu}_l+T^{\mu\nu}_{\gamma\rightarrow u}$, where 
\begin{equation}
T^{\mu\nu}_i=w_iu^\mu u^\nu+g^{\mu\nu}p_i
\end{equation}
for plasma component $i$ ($i=b, l,\gamma\rightarrow u$ ), and $p_i$, $w_i$ are the corresponding pressure and  specific enthalpy.   Since the temperature inside the shock is 
roughly $\Gamma m_ec^2$, the downstream moving leptons are relativistic and we henceforth adopt the approximation $w_l=4p_l=4\n kT$, $w_b=n m_pc^2+4n_ekT$,
neglecting the electron rest mass in the latter expression.
The energy of counterstreaming photons, on the other hand, is typically $m_ec^2$, so that inside the shock, where $\Gamma>>1$, their contribution to energy and momentum balance
can be neglected.   Under this approximation the net energy flux can be expressed as, 
\begin{equation}
T^{0x}=\Gamma^2\beta [m_pc^2n+4(n_e+\n)m_ec^2\hat{T}]= m_pc^2 n \Gamma^2\beta(1+(\x+1)\mu \hat{T}),
\end{equation}
in terms of the mass ratio $\mu=m_e/m_p$ and the dimensionless temperature $\hat{T}=kT/m_ec^2$.  Energy conservation, $\partial_\mu T^{0\mu}=0$, the boundary conditions $\x(z\rightarrow\infty)=\hat{T}(z\rightarrow\infty)=0$, $\Gamma(z\rightarrow\infty)=\Gamma_u$, and baryon number conservation, $n\Gamma\beta=n_u\Gamma_u\beta_u$, then yield the conservation law:
\begin{equation}
\Gamma (1+(\x+1)\mu \hat{T})=\Gamma_u.\label{app:Gamma-cons}
\end{equation}

\bibliographystyle{mnras}

\begin{thebibliography}{}
\makeatletter
\relax
\def\mn@urlcharsother{\let\do\@makeother \do\$\do\&\do\#\do\^\do\_\do\%\do\~}
\def\mn@doi{\begingroup\mn@urlcharsother \@ifnextchar [ {\mn@doi@}
  {\mn@doi@[]}}
\def\mn@doi@[#1]#2{\def\@tempa{#1}\ifx\@tempa\@empty \href
  {http://dx.doi.org/#2} {doi:#2}\else \href {http://dx.doi.org/#2} {#1}\fi
  \endgroup}
\def\mn@eprint#1#2{\mn@eprint@#1:#2::\@nil}
\def\mn@eprint@arXiv#1{\href {http://arxiv.org/abs/#1} {{\tt arXiv:#1}}}
\def\mn@eprint@dblp#1{\href {http://dblp.uni-trier.de/rec/bibtex/#1.xml}
  {dblp:#1}}
\def\mn@eprint@#1:#2:#3:#4\@nil{\def\@tempa {#1}\def\@tempb {#2}\def\@tempc
  {#3}\ifx \@tempc \@empty \let \@tempc \@tempb \let \@tempb \@tempa \fi \ifx
  \@tempb \@empty \def\@tempb {arXiv}\fi \@ifundefined
  {mn@eprint@\@tempb}{\@tempb:\@tempc}{\expandafter \expandafter \csname
  mn@eprint@\@tempb\endcsname \expandafter{\@tempc}}}

\bibitem[\protect\citeauthoryear{{Balberg} \& {Loeb}}{{Balberg} \&
  {Loeb}}{2011}]{balberg2011}
{Balberg} S.,  {Loeb} A.,  2011, \mn@doi [\mnras]
  {10.1111/j.1365-2966.2011.18505.x}, \href
  {http://adsabs.harvard.edu/abs/2011MNRAS.414.1715B} {414, 1715}

\bibitem[\protect\citeauthoryear{{Barniol Duran}, {Nakar}, {Piran}  \&
  {Sari}}{{Barniol Duran} et~al.}{2015}]{barniol-duran2015}
{Barniol Duran} R.,  {Nakar} E.,  {Piran} T.,   {Sari} R.,  2015, \mn@doi
  [\mnras] {10.1093/mnras/stv011}, \href
  {http://adsabs.harvard.edu/abs/2015MNRAS.448..417B} {448, 417}

\bibitem[\protect\citeauthoryear{{Bromberg}, {Mikolitzky}  \&
  {Levinson}}{{Bromberg} et~al.}{2011}]{bromberg2011}
{Bromberg} O.,  {Mikolitzky} Z.,   {Levinson} A.,  2011, \mn@doi [\apj]
  {10.1088/0004-637X/733/2/85}, \href
  {http://adsabs.harvard.edu/abs/2011ApJ...733...85B} {733, 85}

\bibitem[\protect\citeauthoryear{{Bromberg}, {Nakar}, {Piran}  \&
  {Sari}}{{Bromberg} et~al.}{2012}]{bromberg2012}
{Bromberg} O.,  {Nakar} E.,  {Piran} T.,   {Sari} R.,  2012, \mn@doi [\apj]
  {10.1088/0004-637X/749/2/110}, \href
  {http://adsabs.harvard.edu/abs/2012ApJ...749..110B} {749, 110}

\bibitem[\protect\citeauthoryear{{Budnik}, {Katz}, {Sagiv}  \&
  {Waxman}}{{Budnik} et~al.}{2010}]{budnik2010}
{Budnik} R.,  {Katz} B.,  {Sagiv} A.,   {Waxman} E.,  2010, \mn@doi [\apj]
  {10.1088/0004-637X/725/1/63}, \href
  {http://adsabs.harvard.edu/abs/2010ApJ...725...63B} {725, 63}

\bibitem[\protect\citeauthoryear{{Chevalier} \& {Irwin}}{{Chevalier} \&
  {Irwin}}{2011}]{chevalier2011}
{Chevalier} R.~A.,  {Irwin} C.~M.,  2011, \mn@doi [\apjl]
  {10.1088/2041-8205/729/1/L6}, \href
  {http://adsabs.harvard.edu/abs/2011ApJ...729L...6C} {729, L6}

\bibitem[\protect\citeauthoryear{{Johnson} \& {McKee}}{{Johnson} \&
  {McKee}}{1971}]{johnson1971}
{Johnson} M.~H.,  {McKee} C.~F.,  1971, \mn@doi [\prd]
  {10.1103/PhysRevD.3.858}, \href
  {http://adsabs.harvard.edu/abs/1971PhRvD...3..858J} {3, 858}

\bibitem[\protect\citeauthoryear{{Lazzati}, {Morsony}  \& {Begelman}}{{Lazzati}
  et~al.}{2010}]{lazzati2010}
{Lazzati} D.,  {Morsony} B.~J.,   {Begelman} M.~C.,  2010, \mn@doi [\apj]
  {10.1088/0004-637X/717/1/239}, \href
  {http://adsabs.harvard.edu/abs/2010ApJ...717..239L} {717, 239}

\bibitem[\protect\citeauthoryear{{Matzner}, {Levin}  \& {Ro}}{{Matzner}
  et~al.}{2013}]{matzner2013}
{Matzner} C.~D.,  {Levin} Y.,   {Ro} S.,  2013, \mn@doi [\apj]
  {10.1088/0004-637X/779/1/60}, \href
  {http://adsabs.harvard.edu/abs/2013ApJ...779...60M} {779, 60}


\bibitem[\protect\citeauthoryear{{Nakar} \& {Sari}}{{Nakar} \&
  {Sari}}{2012}]{nakar2012}
{Nakar} E.,  {Sari} R.,  2012, \mn@doi [\apj] {10.1088/0004-637X/747/2/88},
  \href {http://adsabs.harvard.edu/abs/2012ApJ...747...88N} {747, 88}


\bibitem[\protect\citeauthoryear{{Nakar}}{{Nakar}}{2015}]{nakar2015}
{Nakar} E.,  2015, \mn@doi [\apj] {10.1088/0004-637X/807/2/172}, \href
  {http://adsabs.harvard.edu/abs/2015ApJ...807..172N} {807, 172}

\bibitem[\protect\citeauthoryear{{Nakar} \& {Piran}}{{Nakar} \&
  {Piran}}{2017}]{nakar2017}
{Nakar} E.,  {Piran} T.,  2017, \mn@doi [\apj] {10.3847/1538-4357/834/1/28},
  \href {http://adsabs.harvard.edu/abs/2017ApJ...834...28N} {834, 28}

\bibitem[\protect\citeauthoryear{{Ofek} et~al.,}{{Ofek}
  et~al.}{2010}]{ofek2010}
{Ofek} E.~O.,  et~al., 2010, \mn@doi [\apj] {10.1088/0004-637X/724/2/1396},
  \href {http://adsabs.harvard.edu/abs/2010ApJ...724.1396O} {724, 1396}

\bibitem[\protect\citeauthoryear{{Pan} \& {Sari}}{{Pan} \&
  {Sari}}{2006}]{pan2006}
{Pan} M.,  {Sari} R.,  2006, \mn@doi [\apj] {10.1086/502958}, \href
  {http://adsabs.harvard.edu/abs/2006ApJ...643..416P} {643, 416}

\bibitem[\protect\citeauthoryear{{Piran}, {Nakar}, {Mazzali}  \&
  {Pian}}{{Piran} et~al.}{2017}]{piran2017}
{Piran} T.,  {Nakar} E.,  {Mazzali} P.,   {Pian} E.,  2017, preprint, \href
  {http://adsabs.harvard.edu/abs/2017arXiv170408298P} {} (\mn@eprint {arXiv}
  {1704.08298})

\bibitem[\protect\citeauthoryear{Ramirez-Ruiz, Celotti  \& Rees}{Ramirez-Ruiz
  et~al.}{2002}]{ramirez-ruiz2002}
Ramirez-Ruiz E.,  Celotti A.,   Rees M.~J.,  2002, \mn@doi [Mon. Not. R.
  Astron. Soc. Vol. 337, Issue 4, pp. 1349-1356.]
  {10.1046/j.1365-8711.2002.05995.x}, 337, 1349

\bibitem[\protect\citeauthoryear{{Salbi}, {Matzner}, {Ro}  \& {Levin}}{{Salbi}
  et~al.}{2014}]{salbi2014}
{Salbi} P.,  {Matzner} C.~D.,  {Ro} S.,   {Levin} Y.,  2014, \mn@doi [\apj]
  {10.1088/0004-637X/790/1/71}, \href
  {http://adsabs.harvard.edu/abs/2014ApJ...790...71S} {790, 71}

\bibitem[\protect\citeauthoryear{{Svirski} \& {Nakar}}{{Svirski} \&
  {Nakar}}{2014a}]{svirski2014}
{Svirski} G.,  {Nakar} E.,  2014a, \mn@doi [\apj]
  {10.1088/0004-637X/788/2/113}, \href
  {http://adsabs.harvard.edu/abs/2014ApJ...788..113S} {788, 113}

\bibitem[\protect\citeauthoryear{{Svirski} \& {Nakar}}{{Svirski} \&
  {Nakar}}{2014b}]{svirski2014a}
{Svirski} G.,  {Nakar} E.,  2014b, \mn@doi [\apjl]
  {10.1088/2041-8205/788/1/L14}, \href
  {http://adsabs.harvard.edu/abs/2014ApJ...788L..14S} {788, L14}

\bibitem[\protect\citeauthoryear{{Tan}, {Matzner}  \& {McKee}}{{Tan}
  et~al.}{2001}]{tan2001}
{Tan} J.~C.,  {Matzner} C.~D.,   {McKee} C.~F.,  2001, \mn@doi [\apj]
  {10.1086/320245}, \href {http://adsabs.harvard.edu/abs/2001ApJ...551..946T}
  {551, 946}

\makeatother
\end{thebibliography}

\end{document}